\documentclass{article}
\usepackage{spconf,amsmath,graphicx}
\usepackage{booktabs}
\usepackage{color,amsmath,url,times, tabularx,bbm,amssymb,multirow}

\title{Improving Weakly Supervised Sound Event Detection with Causal Intervention}

\name{Yifei Xin$^{1}$,
      Dongchao Yang$^{1}$,
      Fan Cui$^{2}$, 
      Yujun Wang$^{2}$,
      Yuexian Zou$^{1,*}$\thanks{This paper was partially supported by NSFC (No: 62176008) and Shenzhen Science \& Technology Research Program (No:GXWD20201231165807007-20200814115301001). Special acknowledgements are given to Xiaomi for its support.}\thanks{$^{*}$ Yuexian Zou is the corresponding author.}}
      
\address{
  $^{1}$School of ECE, Peking University, Shenzhen, China\\
  $^{2}$Xiaomi Corporation, Beijing, China}
\begin{document}
%
\maketitle
\begin{abstract}
Existing weakly supervised sound event detection (WSSED) work has not explored both types of co-occurrences simultaneously, i.e., some sound events often co-occur, and their occurrences are usually accompanied by specific background sounds, so they would be inevitably entangled, causing misclassification and biased localization results with only clip-level supervision. To tackle this issue, we first establish a structural causal model (SCM) to reveal that the context is the main cause of co-occurrence confounders that mislead the model to learn spurious correlations between frames and clip-level labels. Based on the causal analysis, we propose a causal intervention (CI) method for WSSED to remove the negative impact of co-occurrence confounders by iteratively accumulating every possible context of each class and then re-projecting the contexts to the frame-level features for making the event boundary clearer. Experiments show that our method effectively improves the performance on multiple datasets and can generalize to various baseline models.
\end{abstract}
\begin{keywords}
Causal intervention, weakly supervised sound event detection, structural causal model
\end{keywords}
\section{Introduction}
\label{sec:intro}
Sound event detection (SED) involves two subtasks: one is to recognize the types of sound events in an audio clip (audio tagging), and the other is to pinpoint their onset and offset times (localization). Since frame-level labels are costly to collect, weakly supervised sound event detection (WSSED) \cite{kong2020sound,xin2022audio} has gained an increasing research interest, which has only access to weak clip-level labels in the training stage, yet requires to perform the frame-level prediction of onset and offset times during evaluation. 

However, one challenging problem of WSSED is that some sound events often co-occur in an audio clip (e.g., the two classes “train” and “train horn” in DCASE2017 task4 dataset \cite{mesaros2017dcase}). As a result, it is difficult to distinguish those frequently co-occurring sound events in an audio recording, since the model will inevitably relate the sound event class of “train” with that of “train horn”, which interferes with the recognition and detection of each other. Some approaches have been proposed to address this issue. In \cite{imoto2019sound}, graph Laplacian regularization was introduced to model the co-occurrence of sound events for strong labeled SED. In \cite{lin2020specialized}, Lin et al. proposed a disentangled feature, which re-models the high-level feature space so that the feature subspace can be different for each sound event. However, the co-occurrence issue is not just between sound events as sound events also usually co-occur with specific background sounds. Thus, the sound events and background sounds would be inevitably entangled, causing the model to falsely generate ambiguous frame-level localization results with only clip-level supervision. In this work, we target the co-occurrence issue from the two aspects mentioned above for WSSED, which we call “entangled context”. The “entangled context” would lead to misclassification and biased localization results, including the wrongly confusing co-occurring sound events and entangled background sounds. Therefore, we argue that resolving the “entangled context” issue is essential for WSSED. 

In this paper, we attempt to address this issue with causal intervention (CI) method, called CI-WSSED. CI-WSSED attributes the “entangled context” to the frequently co-occurring sound events and specific background sounds that mislead the model to learn spurious correlations between frames and clip-level labels. To find those frames which truly contribute to the clip-level labels in an audio clip, we first establish a structural causal model (SCM) \cite{galanti2020critical} to clarify the causal relation among frame-level features, contexts, and clip-level labels, revealing that the context is the main cause of co-occurrence confounders. Ideally, if we could collect enough audio clips covering all the combinations of different sound event co-occurrences under various background sounds in a balanced distribution, we can distinguish any sound event from them easily. However, it is labor-intensive or even impossible to collect such a huge dataset for each sound event class. To this end, we employ causal intervention to intervene the input to be under any possible context in an approximate way. Based on the causal analysis, CI-WSSED is then designed to operate in an iterative procedure, which is achieved by accumulating the contextual information for each class and then employing it as attention to enhance the frame-level representation for making the sound event boundary clearer. 

In summary, the contributions of this paper are as follows: 1) Our work is the first to concern and reveal the “entangled context” issue of WSSED from both aspects of entangled co-occurring sound events and background sounds. 2) We are the first to introduce causal intervention into WSSED for the “entangled context” issue and design a new network structure, called CI-WSSED, to embed the causal intervention into the WSSED pipeline with an end-to-end scheme. 3) Experiments show that our CI-WSSED yields significant performance gains on WSSED datasets and can generalize to various baselines.

\begin{figure}[t]
  \centering
  \includegraphics[width=1.0\linewidth]{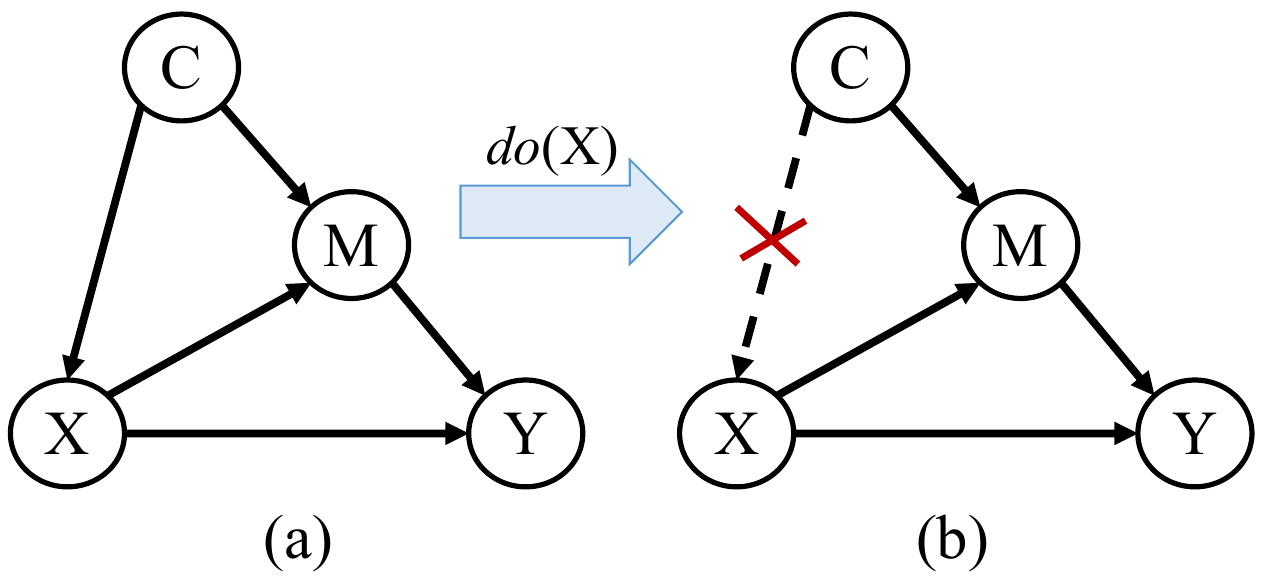}
  \caption{(a) The structural causal model (SCM) for WSSED. (b) The intervened SCM based on backdoor adjustment for WSSED.}
  \label{fig:adapt}
  \vspace*{-\baselineskip}
\end{figure}

\section{Causal Intervention}
\label{sec:c}
The goal of causal learning is to enable the model to pursue causal effects: it can eliminate the spurious bias and disentangle the desired model effects by pursuing the true causal effect \cite{qin2021causal}. Nowadays, the structural causal model (SCM) is commonly utilized in causal learning scenarios. SCM employs a graphical formalism in which nodes are represented as random variables and directed edges represent the direct causal relationship between these variables. As shown in Fig. 1(a), the conditional distribution $P(Y|X)$ expresses the likelihood of $Y$ given $X$, where $Y$ is not only caused by $X$ via $X \rightarrow Y$, but also $C$ via the correlation $C \rightarrow X \rightarrow Y$. 

To find the causal effect of variable $X$ on variable $Y$, do-calculus is introduced \cite{glymour2016causal}. In detail, causal intervention \cite{zhang2020causal} fixes the target variable $X$ to a constant $x$, denoted as $do(X = x)$, rendering it independent of its causes, so the causal effect of $X$ on $Y$ is formulated as:
\begin{equation}
    P(Y|do(X = x)),
\label{eq:crma}
\end{equation}
where the do-notation denotes intervening to set variable $X$ to the value $x$, thereby removing all incoming arrows to the variable $X$, as shown in Fig. 1(b). In this way, we can leverage $P(Y|do(X))$ to pursue the true causality and remove the negative effect of confounders. 

A straightforward way to intervene $X$ is conducting a randomized controlled trial \cite{chalmers1981method} with an ideal huge dataset, which contains audio clips of all kinds of sound event co-occurrences under various background sounds. Therefore, the spurious correlation $C \rightarrow X$ is cutoff and then $P(Y|X) = P(Y|do(X))$. Since this kind of intervention is impossible due to the huge cost of collecting such a large dataset for each sound event, we apply the backdoor adjustment \cite{adib2020causally} to approximate it. In the following section, we will detail how we leverage this solution to WSSED in our approach.

\begin{figure}[t]
  \centering
  \includegraphics[width=1.0\linewidth]{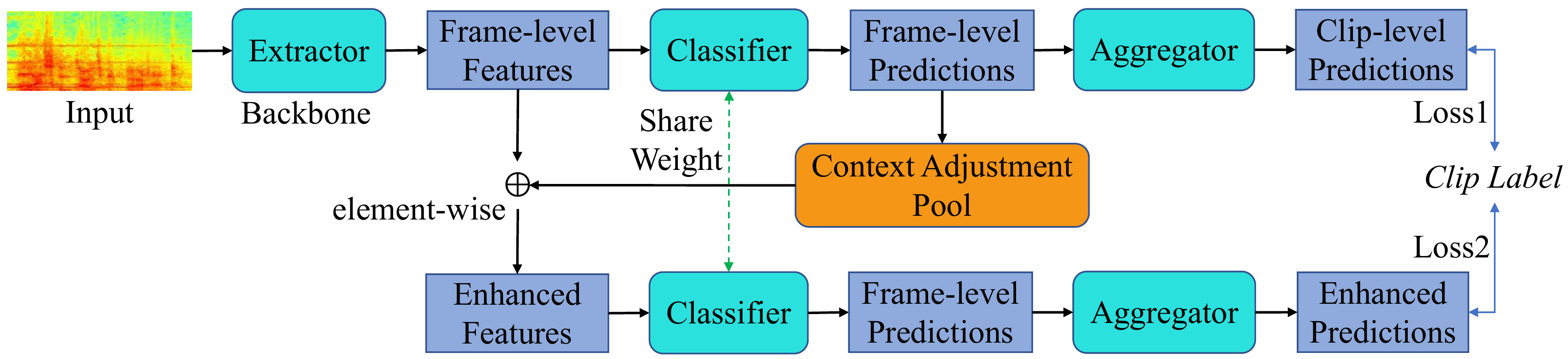}
  \caption{Overview of our proposed CI-WSSED approach.}
  \label{fig:adapt-pic}
  \vspace*{-\baselineskip}
\end{figure}
\section{Proposed Method}
\label{sec:cns}
\begin{table*}
  \caption{Performance comparison of CI-WSSED and baseline models on the DCASE2017 task4 validation and evaluation set.}
  \centering
  \label{tab:freq}
  \begin{tabular}{l|ccc|ccccc}
    \toprule
    \multirow{3}{*}{Method} & \multicolumn{3}{c|}{Validation Set} & \multicolumn{5}{c}{Evaluation Set} \\
    & AT-mAP & AT-F1 & SED-mAP & AT-mAP & AT-F1 & SED-mAP & Seg-F1 & Event-F1\\
    \midrule
    Winner SED \cite{lee2017ensemble} & - & - & - & - & 0.526 & - & 0.555 & -\\
    CDur \cite{dinkel2021towards} & - & - & - & - & 0.553 & - & 0.508 & 0.152\\
    CNN-biGRU \cite{kong2020sound} & 0.650 & 0.555 & 0.456 & 0.650 & 0.632 & 0.444 & 0.564 & -\\
    CNN-Transformer \cite{kong2020sound} & 0.653 & 0.557 & 0.437 & 0.656 & 0.629 & 0.454 & 0.556 & -\\
    HTSAT \cite{chen2022hts} & 0.661 & 0.560 & 0.524 & 0.668 & 0.636 & 0.535 & 0.587 & 0.178\\
    \midrule
    CDur-CI & - & - & - & - & \textbf{0.561} & - & \textbf{0.511} & \textbf{0.164}\\
    CNN-biGRU-CI & \textbf{0.661} & \textbf{0.566} & \textbf{0.462} & \textbf{0.662} & \textbf{0.641} & \textbf{0.453} & \textbf{0.570} & -\\
    CNN-Transformer-CI & \textbf{0.662} & \textbf{0.568} & \textbf{0.449} & \textbf{0.666} & \textbf{0.637} & \textbf{0.461} & \textbf{0.561} & -\\
    HTSAT-CI & \textbf{0.672} & \textbf{0.572} & \textbf{0.531} & \textbf{0.678} & \textbf{0.644} & \textbf{0.544} & \textbf{0.592} & \textbf{0.191}\\
  \bottomrule
\end{tabular}
\vspace*{-\baselineskip}
\end{table*}
\subsection{Structural Causal Model for WSSED}
\label{subsec:net}
In this part, we will explain why the entangled context hinders the sound event classification and localization performance in WSSED. We formulate the causalities among frame-level features $X$, contexts $C$, frame-level predictions $M$, and predicted clip-level labels $Y$, with a structural causal model (SCM) illustrated in Fig. 1(a). The direct links represent the causalities between the two nodes: cause → effect.

$C \rightarrow X$: We denote $C$ as the prior context knowledge. This link represents that the extractor produces frame-level features $X$ under the effect of contexts $C$. Although the contextual information improves the association between the frame-level features $X$ and predicted labels $Y$ via $P(Y|X)$, $P(Y|X)$ mistakenly associates non-causal but positively correlated frames to labels. There are many possible sources of the contextual bias, which may be from the training process (e.g., batch normalization) and biased datasets. For example, the sound events of “train horn” and “train” often occur in co-occurrence with each other, so the model would be confounded to establish a spurious correction between the two sound event classes, and this causal link will exacerbate the existing bias. 

$C \rightarrow M \leftarrow X$: We term the mediator $M$ as the $X$-specific context, which is directly from $X$ but essentially inherited from $C$. Specifically, the context is a combination of various other sound events, for instance, when multiple sound events (e.g., “children playing”, “street music”, and “dog bark”) occur in an audio clip, the “children playing” can be seen as the label with its context including “street music” and “dog bark”, and the same holds true when the lead is “street music” or “dog bark”. 

$X \rightarrow Y \leftarrow M$: These links denote that the sound event itself and its context together affect the final prediction. However, a general $C$ cannot directly influence the predicted clip-level labels $Y$. Thus, in addition to the direct effect $X \rightarrow Y$, $Y$ is also the effect of the $X$-specific context $M$, which contains the timestamp information of the sound event and its context.

Considering the impact of contexts $C$ on frame-level features $X$, we will cut off the link from $C$ to $X$. Next, we will introduce a causal intervention method to mitigate the negative impact of contextual bias. 

\subsection{Causal Intervention via Backdoor Adjustment}
\label{subsec:recs}
As shown in Fig. 1(b), we propose to use $P(Y|do(X))$ based on the backdoor adjustment \cite{zhang2020causal,shao2021improving} to remove the context confounder and pursue the true causality between $X$ and $Y$. The key idea is to cut off the link $C \rightarrow X$, and stratify $C$ into pieces $C= \left\{c_1, c_2, . . . , c_k\right\}$, where $c_i$ denotes the $i^{th}$ class context. Formally, we have
\begin{equation}
    P(Y|do(X)) = \sum^{k}_{i} P(Y|X = x, M = f(x, c_i))P(c_i),
\label{eq:crm}
\end{equation}
where $f(x, c_i)$ represents that $M$ is formed by the combination of $X$ and $C$, and $k$ is the number of sound event classes. As $C$ is no longer correlated with $X$, the causal intervention guarantees $X$ to have an equal chance of incorporating every context $c_i$ into $Y$’s prediction, based on the proportion of each $c_i$ in the whole. However, the cost of the network forward propagation for all the $k$ classes is expensive. Thanks to the Normalized Weighted Geometric Mean \cite{ding2019neural}, we can optimize Eq. (2) to approximate the above expectation by moving the outer sum ${\sum}^{n}_{i}P(c_i)$ into the feature level 
\begin{equation}
    P(Y|do(X)) \approx P(Y|X = x, M = \sum^{k}_{i}f(x, c_i)P(c_i)).
\label{eq:crm}
\end{equation}
Thus, we only need to feed-forward the network once instead of $k$ times. To simplify the formula, we assume roughly the same number of samples for each class in the dataset, so $P(c_i)$ is set as the uniform $1/k$. After simplifying Eq. (3), we have 
\begin{equation}
    P(Y|do(X)) \approx P(Y|x \oplus \frac{1}{k}\sum^{k}_{i}f(x, c_i)),
\label{eq:crm}
\end{equation}
where $\oplus$ denotes projection. Therefore, the “entangled context” issue has been transferred to the calculation of $\sum^{k}_{i}f(x, c_i)$. Next, we will introduce a context adjustment pool to represent $\sum^{k}_{i}f(x, c_i)$.

\subsection{Network Structure}
\label{subsec:re}
In this part, we implement causal intervention for WSSED with a novel network structure, called CI-WSSED. Fig. 2 illustrates the overview of our CI-WSSED. First, the feature extractor (e.g., the CNN, RNN or Transformer-based backbone) takes the mel-spectrogram as input and produces high-level features $X \in \mathbb{R}^{c \times n}$, where $c$ is the number of channels and $n$ is the number of audio frames. Then, the frame-level features $X$ are fed into the classifier with a fully connected layer followed by an aggregator (e.g., an average pooling function) to produce clip-level prediction results.

We maintain a context adjustment pool $Q \in \mathbb{R}^{k \times n}$ for each sound event class during the training stage, which is the core of our CI-WSSED. According to Eq. (4), $Q$ is designed to continuously store the contextual information of each occurring sound event, and then re-project the accumulated contexts onto the frame-level features $X$ generated by the backbone to produce enhanced features $X^e \in \mathbb{R}^{c \times n}$. In detail, the context adjustment pool is updated by fusing the contexts of each occurring sound event in frame-level predictions $M=f(x, c_i)$, which is followed by a batch normalization: 
\begin{equation}
    Q_j = BN(Q_j + \lambda \times M_j),
\label{eq:crm}
\end{equation}
where $j$ represents the class index of each occurring sound event that we can get from the clip-level labels of each audio clip, and $\lambda$ denotes the update rate. Then, the enhancement of frame-level features can be formulated as:
\begin{equation}
    X^e = X + X \odot Conv(Q_{j}),
\label{eq:crm}
\end{equation}
where $\odot$ denotes matrix dot product and $Conv$ represents the $1 \times 1$ convolution. In this way, we can not only mitigate the impact of “entangled context” including entangled co-occurring sound events and background sounds, but also spotlight the active regions of the frame-level features, thus reducing classification errors and boosting localization performance.

During the training phase, our proposed network learns to minimize the cross-entropy losses for both classification branches. Specifically, we adopt two classifiers with shared weights for the two branches. The first classifier is used to produce initial prediction scores $S = \left\{s_1, s_2, ..., s_k\right\}$, and the second classifier is accountable for generating more accurate prediction scores $S^e = \left\{s^e_1, s^e_2, ..., s^e_k\right\}$ using the enhanced frame-level features. Then, the cross-entropy losses for both classification branches are optimized to train the two classifiers together in an end-to-end pipeline. The overall loss function $L$ is formulated below:
\begin{equation}
    L = (-\sum^{k}_{i=1}s^*_ilog(s_i)) + (-\sum^{k}_{i=1}s^*_ilog(s_i^e)),
\label{eq:crm}
\end{equation}
where $s^*$ is the ground-truth label of an audio clip. While in the inference stage, we use the enhanced features to produce the final frame-level prediction results.

\section{Experiments and Results}
\label{sec:exp}
\subsection{Dataset}
We evaluate our method on the two public sound event detection datasets: DCASE2017 task4 \cite{mesaros2017dcase} and UrbanSED \cite{salamon2017scaper} datasets. The DCASE2017 task4 – Large-scale weakly supervised sound event detection for smart cars dataset is comprised of a training subset with 51172 audio clips, a validation subset with 488 audio clips, and an evaluation set with 1103 audio clips, including 17 sound events. The UrbanSED dataset has 10 event labels within urban environments, divided into 6000 training, 2000 validation, and 2000 evaluation audio clips. 
\begin{table}
  \caption{Performance comparison of CI-WSSED and baseline models on the weakly labelled UrbanSED test set.}
  \centering
  \label{tab:freq}
  \begin{tabular}{cccl}
    \toprule
    Method & AT-F1 & Seg-F1 & Event-F1\\
    \midrule
    Base-CNN \cite{salamon2017scaper} & - & 0.560 & -\\
    HTSAT \cite{chen2022hts} & 0.771 & 0.644 & 0.210\\
    CDur \cite{dinkel2021towards} & 0.771  & 0.647 & 0.217\\
    \midrule
    HTSAT-CI & \textbf{0.776} & \textbf{0.646} & \textbf{0.216}\\
    CDur-CI & \textbf{0.774} & \textbf{0.648} & \textbf{0.220}\\
  \bottomrule
\end{tabular}
\vspace*{-\baselineskip}
\end{table}
\subsection{Baseline Models and Training Details}
To evaluate the effectiveness and generalization of our CI-WSSED, we apply our method to multiple baseline systems, including CDur \cite{dinkel2021towards}, CNN-biGRU \cite{kong2020sound}, CNN-Transformer \cite{kong2020sound} and HTSAT \cite{chen2022hts}. CDur consists of a 5-layer CNN and a bidirectional Gated Recurrent Unit (biGRU), while the CNN-biGRU system is modeled by a 9-layer CNN with a biGRU. The CNN-Transformer consists of a 9-layer CNN with one transformer block. HTSAT uses the Swin Transformer \cite{liu2021swin} backbone with ImageNet-pretraining, where we use 3 network groups with 2, 2, 6 swin-transformer blocks for the DCASE2017 task4 dataset, and for the UrbanSED dataset, we only adopt two stages with 2, 2 swin-transformer blocks. The update rate is set as $\lambda$ = 0.01. We use audio tagging mAP (AT-mAP), audio tagging F1 score (AT-F1), sound event detection mAP (SED-mAP), Segment-F1 score (Seg-F1) and Event-F1 score to evaluate our method.

\subsection{Results on DCASE2017 Task4}
We first report our experiment results on both DCASE2017 task4 validation set and evaluation set in Table 1. We use baseline-CI to represent baseline models using our causal intervention method. It can be seen that our method achieves significant performance boosts on all baseline models, especially on the mAP, AT-F1 and Event-F1 metrics, which demonstrates the effectiveness of our CI-WSSED in reducing classification errors and localizing entire sound events.

\subsection{Results on UrbanSED}
As shown in Table 2, we compare previous methods on the UrbanSED dataset with our CI-WSSED. It is clear that our CI-WSSED also achieves consistent improvements compared to the previous corresponding models. Notably, after applying the causal intervention method, the performance gain of UrbanSED is not as significant as that of the DCASE2017 task4 dataset. We infer the reason is that there are many sound event classes within the DCASE2017 task4 dataset that often co-occur \cite{dinkel2021towards}, such as the sound events of “train” and “train horn”, as well as “car”, “car alarm”, and “car passing by”, so the DCASE2017 task4 dataset suffers more from the “entangled context” and thus benefits more from our CI-WSSED.

\section{Conclusions}
\label{sec:conclusion}
In this paper, we target the “entangled context” problem in the WSSED task from both aspects of entangled co-occurring sound events and background sounds. Through analyzing the causal relationship between frame-level features, contexts, and clip-level labels with the help of the SCM, we pinpoint the context as a co-occurrence confounder and then propose an end-to-end CI-WSSED method to deal with the effect. Experiments show that the “entangled context” is a practical issue within the WSSED task and our CI-WSSED pipeline can effectively boost the performance of WSSED on multiple datasets and generalize to various baseline models.

\bibliographystyle{IEEE.bst}
\bibliography{refs.bib}

\end{document}